\documentclass[11pt]{article}
\usepackage{acl}
\usepackage{times}
\usepackage{latexsym}
\usepackage{amsmath}
\usepackage{amssymb}
\usepackage{booktabs}
\usepackage{multirow}
\usepackage{graphicx}
\usepackage{tikz}
\usetikzlibrary{arrows.meta, positioning, shapes.geometric}
\usepackage{xcolor}
\usepackage{microtype}
\usepackage{hyperref}
\usepackage{algorithm}
\usepackage{algpseudocode}
\usepackage{enumitem}

\newcommand{\sepo}{\textsc{Sepo}}
\newcommand{\Jsepo}{J_{\text{SEPO}}}

\title{Safe Equilibrium Policy Optimization for Strategic Agent Policies}

\author{
  Karthika Arumugam$^{*\dagger}$ \quad Kiran Kumar Manku$^{*\dagger}$ \quad Amit Dhanda$^{*\dagger}$ \\[4pt]
  $^\dagger$\textit{Amazon, USA} \\
  $^*$\textit{Equal contribution}
  \thanks{This work does not relate to the author's position at Amazon.}
}

\begin{document}
\maketitle
\begin{abstract}
Language models fine-tuned with reinforcement learning typically optimize for task reward, ignoring multi-agent strategic structure. Because these agents condition on natural language game-state descriptions and emit actions through free-form generation, strategic failure modes --- exploiting weaker opponents, coordinating on harmful equilibria, and externalizing costs are inseparable from the language interface itself. We propose Safe Equilibrium Policy Optimization (\sepo{}), a training objective that augments expected payoff with explicit penalties for exploitability, collusion risk, and externality cost. We implement \sepo{} as a reward signal for Group Relative Policy Optimization (GRPO), applied to Gemma~4 E4B-it and Qwen~3.5-4B after supervised fine-tuning (SFT). Evaluated across five strategic domains: Iterated Prisoner's Dilemma, repeated auctions, two negotiation variants, and Kuhn Poker.  \sepo{} achieves zero exploit-pool advantage in Kuhn Poker for both models, outperforms the base model on safety in four domains, and corrects the over-cooperative behavior introduced by SFT. In negotiation, \sepo{} achieves a positive-safety outcome and only the positive normalized relative advantage of any negotiation configuration. Ablation experiments confirm that per-rollout exploit computation is necessary: a shared constant penalty cancels in GRPO advantage normalization (constant control-variate property), producing zero gradient. To support further research in strategic safety for agents, we release our \href{https://anonymous.4open.science/r/sepo-2668/README.md}{code} and SFT datasets.
\end{abstract}

\section{Introduction}
\label{sec:intro}

\begin{figure*}[t]
  \centering
  \includegraphics[width=0.48\textwidth]{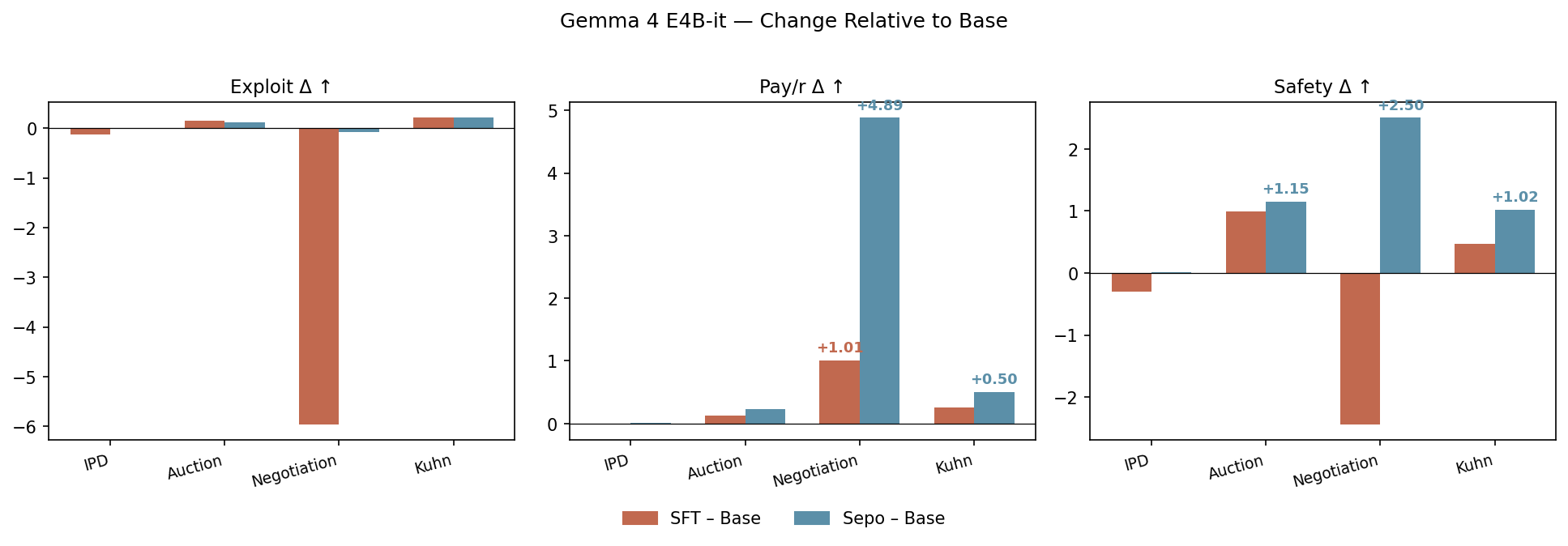}
  \hfill
  \includegraphics[width=0.48\textwidth]{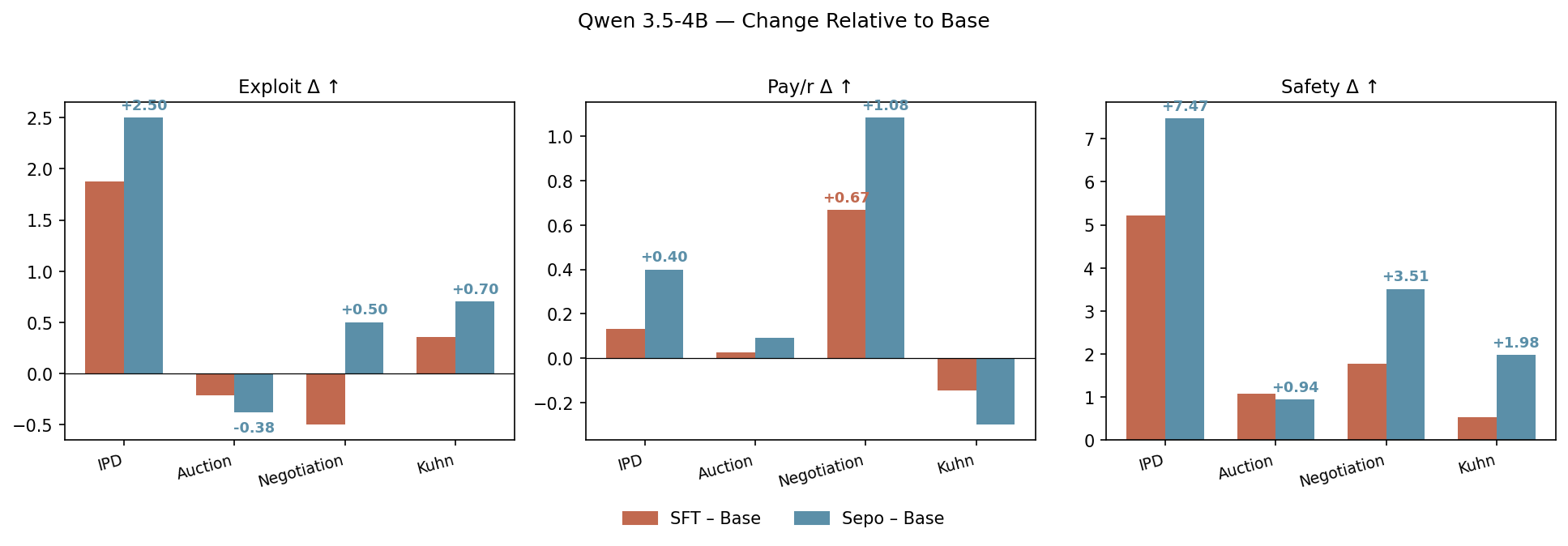}
  \caption{Safety improvement ($\Delta$~Safety = SEPO $-$ Base) across all games. Left: Gemma~4. Right: Qwen 3.5-4B. Positive bars indicate \sepo{} improves over base.}
  \label{fig:delta_overview}
\end{figure*}

Two LLM-powered procurement agents, each serving a different organization, interact repeatedly on a digital marketplace. Without explicit coordination, they converge on a price-splitting pattern that benefits both principals but inflates costs for downstream consumers. No agent was programmed to collude; each maximized its own reward in a repeated game. This is algorithmic collusion --- emergent, hard to detect, and a direct consequence of reward maximization in a strategic setting. As language agents take on higher-stakes roles in negotiation, resource management, automated trading, and multi-party planning, such failures become systemic risks rather than theoretical curiosities \citep{bakhtin2022human,park2023generative,fu2023improving}.

Standard alignment methods address this problem at the wrong level. RLHF \citep{christiano2017deep,ouyang2022training} and Constitutional AI \citep{bai2022constitutional} align a single agent to a fixed preference model, but in multi-agent environments the effective reward function depends on what other agents do. An agent trained to be helpful and harmless in isolation can still learn to exploit weaker opponents, sustain collusive equilibria with cooperative partners, or impose external costs on the shared system. These are not alignment failures in the conventional sense; they are failures of objective design for strategic settings.

We propose Safe Equilibrium Policy Optimization (\sepo{}), a training framework that addresses this gap by augmenting task utility with three penalty terms:
\begin{equation}
\label{eq:sepo}
\Jsepo(\pi) = u(\pi) - \lambda_e \cdot e(\pi) - \lambda_c \cdot c(\pi) - \lambda_x \cdot x(\pi)
\end{equation}
where $u(\pi)$ is expected payoff, $e(\pi)$ is an exploitability proxy estimated against adversarial opponents, $c(\pi)$ is collusion risk estimated against cooperative partners, and $x(\pi)$ is domain-specific externality cost. Each term targets a distinct strategic failure mode; the $\lambda$ coefficients are tuned per-domain.

\sepo{} is implemented as a reward signal applied to language models that generate game actions. Training proceeds in two stages: a supervised fine-tuning (SFT) warm-start on expert strategy traces, followed by \sepo{} optimization computed from fresh opponent interactions at every rollout. We evaluate on Gemma 4 E4B-it and Qwen 3.5-4B across five strategic domains: Iterated Prisoner's Dilemma (IPD), repeated sealed-bid auctions, single-issue negotiation, GTBench multi-issue negotiation \citep{wu2024gtbench}, and Kuhn Poker.

We report four main findings. First, \sepo{} achieves zero exploit-pool advantage in Kuhn Poker for both Gemma 4 and Qwen (at the best checkpoint), converging to the Nash mixed strategy against the adversaries in our pool. Second, in GTBench negotiation --- where private valuations require genuine multi-round inference --- \sepo{} achieves a positive-safety outcome and yields the only positive normalized relative advantage (NRA) of any negotiation configuration. Third, supervised fine-tuning consistently degrades exploit resistance across all games; \sepo{} corrects this regression. Fourth, the per-rollout exploit gradient is the critical implementation choice: a constant exploit penalty cancels in \sepo{}'s advantage normalization and produces zero gradient --- a consequence of the standard constant control-variate property \citep{williams1992simple,kool2019buy}. Computing exploit per-rollout against fresh opponent samples makes the term vary across the group and restores a genuine gradient signal.

Our contributions are:
\begin{enumerate}
\item A training objective for strategic language agents that jointly penalizes exploitability, collusion risk, and externality cost, with formal connections to Nash and correlated equilibrium concepts.
\item An implementation of \sepo{} with per-rollout opponent interaction and a principled three-pool opponent design.
\item An empirical study across five strategic domains demonstrating that \sepo{} improves strategic safety relative to both reward-only RL and SFT baselines, and identifying the conditions under which equilibrium convergence occurs.
\end{enumerate}

\section{Related Work}
\label{sec:related}

\subsection{Language Agents in Strategic Environments}

Language models have been deployed as autonomous agents in strategic domains. \citet{bakhtin2022human} demonstrated human-level Diplomacy play by combining a language model with game-theoretic planning at inference time. \citet{park2023generative} showed that LLM-powered agents produce emergent social behaviors---cooperation, coalition formation, information diffusion---when placed in persistent multi-agent environments. These systems optimize for task completion rather than strategic safety properties of the resulting equilibrium.

Several lines of work study LLMs as players in well-defined games. \citet{fu2023improving} applied self-play with AI feedback to improve LLM negotiation, gaining deal rates and individual payoff. \citet{guo2023suspicion} evaluated GPT-4 with theory-of-mind prompting in imperfect-information games. \citet{mao2025alympics} benchmarked LLMs in classic game-theoretic scenarios, revealing systematic deviations from Nash equilibrium. \citet{zhang2025klevel} proposed $K$-level reasoning to induce higher-order strategic beliefs, improving performance in matrix games. \citet{du2024improving} showed that multi-agent debate improves factual accuracy, suggesting that strategic interaction can serve as an inference-time alignment mechanism.

Foundational NLP work on negotiation predates the current wave of LLM agents. \citet{lewis2017deal} introduced Deal-or-No-Deal and found that reward-maximizing agents developed deceptive strategies. \citet{he2018decoupling} separated strategic planning from language generation for more controllable behavior. \citet{xu2023exploring} studied LLMs in Werewolf, finding that strategic communication emerges from in-context reasoning. These studies collectively establish that objective design shapes strategic behavior in language agents---the central premise of \sepo{}.

\subsection{Multi-Agent RL and Social Dilemmas}

\citet{leibo2017multiagent} introduced sequential social dilemmas, showing that independent RL agents converge to defection even when mutual cooperation yields higher welfare. \citet{lerer2017maintaining} showed that deep RL can maintain cooperation through opponent history conditioning. \citet{foerster2018learning} proposed LOLA, which differentiates through an opponent's anticipated learning step to shape its future policy toward cooperation.

Social preference engineering has been used to modify reward signals. \citet{jaques2019social} used social influence as intrinsic motivation; \citet{hughes2018inequity} embedded inequity aversion into reward functions. \citet{zhang2024scalable} addressed safe MARL with constrained optimization over joint behavior.

\sepo{} shares the goal of safer multi-agent outcomes but decomposes it differently. Where social influence and inequity aversion modify the reward to encourage specific prosocial behaviors, \sepo{} penalizes three failure modes as separate terms. This decomposition supports ablation and per-domain diagnosis.

\subsection{Equilibrium Training and Exploitability Minimization}

Computing or approximating equilibria has driven strategic AI from Go \citep{silver2016mastering} to multiplayer poker \citep{brown2019superhuman}. \citet{brown2019deep} scaled counterfactual regret minimization with neural networks. \citet{heinrich2016deep} proposed Neural Fictitious Self-Play (NFSP) for approximate Nash equilibria in two-player games. \citet{lanctot2017unified} introduced PSRO, iteratively expanding a strategy portfolio via best-response computation. \citet{lockhart2019computing} proposed exploitability descent as a direct gradient-based approach to Nash equilibria.These methods target exploitability alone. \sepo{} incorporates exploitability minimization as one component ($e(\pi)$) and adds collusion risk ($c(\pi)$) and externality cost ($x(\pi)$). A Nash equilibrium can be socially harmful---a cartel is stable but collusive---so equilibrium training alone does not address the full range of strategic safety concerns.

\subsection{Safe and Value-Aligned Optimization}

\citet{christiano2017deep} introduced RLHF by learning reward models from human preferences; \citet{ouyang2022training} scaled this to InstructGPT. \citet{bai2022constitutional} replaced human feedback with AI-generated critiques guided by principles. \citet{rafailov2023direct} eliminated the explicit reward model via direct preference optimization. \citet{achiam2017constrained} introduced CPO for safe single-agent RL with cost constraints. \citet{amodei2016concrete} taxonomized AI safety problems including reward hacking and side effects that intensify in multi-agent settings.

All of these methods assume a single agent in a fixed environment. \sepo{} fills the gap: its penalty terms are computed against other agents' strategies, making safety constraints inherently multi-agent in scope.

\section{Method}
\label{sec:method}

\subsection{The \sepo{} Objective}
\label{sec:objective}

The \sepo{} objective is Equation~\ref{eq:sepo}, repeated here with formal definitions of each term.

\paragraph{Expected payoff $u(\pi)$.} The agent's mean reward against a pool of cooperative training opponents drawn from a domain-specific strategy library. This term recovers the standard task reward objective when used alone.

\paragraph{Exploitability $e(\pi)$.} The mean payoff advantage that adversarial opponents achieve over the agent. We estimate $e(\pi) = \frac{1}{|\mathcal{E}|}\sum_{o \in \mathcal{E}} [u(o, \pi) - u_{\text{fair}}]$, where $\mathcal{E}$ is the exploit pool of adversarial strategy types and $u_{\text{fair}}$ is the payoff the agent would earn against a passive (non-adversarial) opponent (IPD: mutual cooperation payoff; Auction: value-bid payoff; Negotiation: fair-split payoff; Kuhn: ante-only). This estimates exploit-pool advantage rather than true worst-case exploitability. Minimizing $e(\pi)$ pushes the policy toward best-response stability: a policy with $e(\pi) = 0$ cannot be profitably exploited by any opponent in the pool. This approximates Nash equilibrium when the exploit pool spans the adversarial strategy space.

\paragraph{Collusion risk $c(\pi)$.} The rate at which the agent engages in harmful coordination when paired with a collusive partner. We estimate $c(\pi)$ by measuring joint behavior with collusive partner types (e.g., partners that reciprocate high extraction in resource games, or partners that bid low in auctions). Penalizing $c(\pi)$ discourages correlated equilibria that are jointly profitable but socially harmful.

\paragraph{Externality cost $x(\pi)$.} Domain-specific harm imposed outside the bilateral interaction: welfare shortfall from the social optimum in social dilemmas, depletion in resource games, allocative inefficiency in auctions, and breakdown rate in negotiation. Minimizing $x(\pi)$ internalizes external costs that the private payoff function ignores.

\paragraph{Connection to game-theoretic concepts.} The $u(\pi)$ term alone maximizes individual payoff with no equilibrium incentive. Adding $e(\pi)$ drives the policy toward Nash best-response stability. Penalizing $c(\pi)$ discourages privately profitable correlated equilibria that harm the system. Penalizing $x(\pi)$ provides a Pigouvian correction at the policy level. Together, the three terms constitute an objective that simultaneously rewards individual competence and penalizes the three principal modes of strategic failure.

\subsection{\sepo{} Implementation}
\label{sec:grpo}

\sepo{} is trained using Group Relative Policy Optimization \citep[GRPO;][]{shao2024deepseekmath} as the underlying RL algorithm. Throughout the rest of this paper we refer to the full training procedure (GRPO + the \sepo{} objective) simply as \sepo{}. All payoffs are first normalized to mutual-cooperation payoff as 3.0 scale(standard IPD scale) via $s = 3.0/\text{game.max\_payoff}$, making $\lambda$ coefficients transferable across games with different payoff magnitudes; After normalization, the per-rollout scalar reward is:
\begin{equation}
\label{eq:reward}
r_r = u_r - \lambda_e \cdot e - \lambda_c \cdot c - \lambda_x \cdot x
\end{equation}
where $u_r$ is the per-rollout mean normalized payoff, and $e$, $c$, $x$ are the SEPO penalties estimated from auxiliary opponent interactions for that training step (shared across rollouts for the same training opponent).

\paragraph{Opponent pool design.} Three pools serve strictly separated purposes: (1)~\textit{Training pool} --- diverse cooperative strategies that generate the utility signal $u_r$; (2)~\textit{Exploit pool} --- adversarial strategies that punish passive compliance (e.g., AlwaysDefect, AggressiveBid), used to estimate $e(\pi)$; (3)~\textit{Collusive pool} --- cooperative partners that reveal greedy tendencies, used to estimate $c(\pi)$. Each strategy appears in exactly one pool. Assigning the same strategy to multiple pools creates conflicting gradient signals for identical actions in different contexts --- an implementation bug that nullifies the safety gradient.

\paragraph{Per-opponent exploit averaging.} When the exploit pool contains multiple adversaries, pooled averaging lets strong performance against one mask weakness against another. We compute exploitability separately per opponent and then average, ensuring every adversary contributes equally to $e(\pi)$.

\subsection{Policy Gradient Loss}
\label{sec:pg_loss}

\paragraph{Per-round advantage normalization.}
Rather than assigning a single advantage to an entire episode, we normalize advantages \emph{per game round} across the $n$ rollouts for a given training opponent. For round $t$ and rollout $r$:
\begin{align}
\text{reward}_{t,r} &= u_{t,r} - \underbrace{(\lambda_e e + \lambda_c c + \lambda_x x)}_{\text{SEPO penalty, episode-level}} \label{eq:round_reward}\\
A_{t,r} &= \frac{\text{reward}_{t,r} - \overline{\text{reward}}_{t}}{\sigma_t + \delta} \label{eq:advantage}
\end{align}
where $\overline{\text{reward}}_t$ and $\sigma_t$ are the mean and standard deviation across rollouts at round $t$, and $\delta = 10^{-8}$ is a numerical stabilizer (distinct from the PPO clip parameter $\epsilon = 0.2$). Round-level normalization ensures that even a single divergent action at one step of one rollout produces a non-zero advantage signal --- critical when the SFT model is near-deterministic and all rollouts produce identical episode-level outcomes.

\paragraph{Why episode-level reward fails.} Episode-level normalization fails for two compounding reasons. First, a constant penalty $P$ is always removed by mean subtraction regardless of its value: $\text{reward}_r - \overline{\text{reward}} = u_r - \overline{u}$ (the penalty cancels). This is the standard constant control-variate property \citep{williams1992simple,kool2019buy}: any penalty that is the same for every rollout in the group acts as a group-level baseline and contributes zero gradient. \sepo{} breaks this by computing the exploit signal per-rollout against a freshly sampled opponent, so $e(\pi_r)$ varies across the group and contributes a genuine gradient. Second, the SFT warm-start is near-deterministic (temperature 0.8 on a well-trained model often yields the same action sequence), collapsing $\sigma \approx 0$ across rollouts and zeroing all advantages. Per-round normalization breaks the first problem (different rounds have different $u_{t,r}$, so the constant $P$ does \emph{not} cancel), and per-step stochastic variation in $a_{t,r}$ prevents the second. We verified this in preliminary experiments: episode-level \sepo{} produced no exploit improvement over 100 steps (exploit $\approx 0.33$ for IPD, matching the base model), while per-round \sepo{} reached exploit 0.312 by step 35 (Appendix~\ref{app:ablation}).

\paragraph{PPO-style clipped surrogate loss \citep{schulman2017proximal}.} Given advantages $A_{t,r}$, the per-step policy gradient loss is:
\begin{equation}
\label{eq:pgloss}
\begin{split}
\mathcal{L}_{\mathrm{PG}} = -\min\!\bigl(&\rho_{t,r} A_{t,r},\\
  &\mathrm{clip}(\rho_{t,r}, 1{-}\epsilon, 1{+}\epsilon)\cdot A_{t,r}\bigr)
\end{split}
\end{equation}
where $\rho_{t,r} = \exp(\log\pi_\theta(a_{t,r}|s_{t,r}) - \log\pi_{\mathrm{old}}(a_{t,r}|s_{t,r}))$ is the importance ratio and $\epsilon = 0.2$. The KL penalty is one-sided to prevent premature convergence:
\begin{equation}
\label{eq:kl}
\mathcal{L}_{\mathrm{KL}} = \max\!\bigl(0,\; \log\pi_\theta(a|s) - \log\pi_{\mathrm{ref}}(a|s)\bigr)
\end{equation}
The full per-step loss is $\mathcal{L} = \mathcal{L}_{\mathrm{PG}} + \beta \cdot \mathcal{L}_{\mathrm{KL}}$, normalized by $|\mathcal{P}_{\mathrm{train}}| \times n \times T$ and backpropagated immediately (one computation graph per step, no accumulation, keeping peak VRAM constant).

\paragraph{Memory-efficient log-softmax.} Gemma 4's vocabulary size is 262,144. Computing $\log\pi_\theta$ over the full vocabulary for a 512-token generation requires $262144 \times 512 \times 4\,\text{B} \approx 536\,\text{MB}$ per forward pass. We process generated tokens in chunks of 32, reducing peak VRAM by $16\times$ without affecting numerical results.

\paragraph{LoRA reference model.} In LoRA mode, the reference policy $\pi_{\mathrm{ref}}$ is obtained by disabling the LoRA adapters in-place (via \texttt{model.disable\_adapter()}) rather than loading a separate frozen copy. This saves ${\sim}10\,\text{GB}$ VRAM on a 24\,GB GPU, making the full pipeline runnable on a single A100 or RTX 4090.

\subsection{Action Generation and Parse Recovery}
\label{sec:parse}

\paragraph{Stopping criteria.} We implement a custom \texttt{StoppingCriteria} that halts generation when a valid action token appears on the last non-empty line of the output. The criteria respects chain-of-thought: if the output contains an open \texttt{<think>} tag without a closing \texttt{</think>}, generation continues regardless of what appears in the reasoning trace.

\paragraph{Constrained fallback decode.} When the model's free-form generation does not contain a parseable action (parse failures occur in 3--9\% of rounds depending on game and model), we recover a valid action without regex or heuristics via forced next-token decoding:
\begin{enumerate}[noitemsep]
  \item Append the generated reasoning as an assistant turn.
  \item Add a short elicitation user turn: \textit{``State your final action (\textsc{action}/\textsc{vocab}):''}.
  \item Forward-pass the full prompt (no generation); extract logits at the final position.
  \item Set all logits to $-\infty$ except the first sub-word tokens of each valid action string.
  \item Return the argmax --- always a valid action, no fallback needed.
\end{enumerate}
This procedure requires one additional forward pass (no beam search, no sampling) and preserves the model's reasoning context when determining the action.

\begin{algorithm}[t]
\caption{\sepo{} Training Step}
\label{alg:sepo_grpo}
\begin{algorithmic}[1]
\Require $\pi_\theta$, $\pi_{\mathrm{ref}}$, game $G$, pools $\mathcal{P}_{\mathrm{tr}}, \mathcal{P}_{\mathrm{ex}}, \mathcal{P}_{\mathrm{co}}$, weights $\lambda_e, \lambda_c, \lambda_x$, $\beta, \epsilon, n$
\For{each train opponent $o \in \mathcal{P}_{\mathrm{tr}}$}
  \State Run aux.\ episodes vs.\ $\mathcal{P}_{\mathrm{ex}}$, $\mathcal{P}_{\mathrm{co}}$ \Comment{shared across rollouts}
  \State $P \leftarrow \lambda_e e + \lambda_c c + \lambda_x x$ \Comment{episode-level SEPO penalty}
  \For{rollout $r = 1 \ldots n$}
    \State Sample episode $\tau_r = (s_1, a_1, \ldots, s_T, a_T)$ against $o$
    \State Record per-round payoffs $\{u_{t,r}\}_{t=1}^T$ and $\log\pi_{\mathrm{old}}(a_t|s_t)$
  \EndFor
  \For{round $t = 1 \ldots T$}
    \State $\text{rew}_{t,r} \leftarrow u_{t,r} - P$ for each rollout $r$
    \State $A_{t,r} \leftarrow (\text{rew}_{t,r} - \overline{\text{rew}}_t) / (\sigma_t + \delta)$ \Comment{$\delta = 10^{-8}$, cf.\ Eq.~\ref{eq:advantage}}
    \For{rollout $r = 1 \ldots n$}
      \State $\rho \leftarrow \exp(\log\pi_\theta(a_{t,r}|s_{t,r}) - \log\pi_{\mathrm{old}}(a_{t,r}|s_{t,r}))$
      \State $\mathcal{L} \leftarrow -\min(\rho A_{t,r},\, \mathrm{clip}(\rho, 1{-}\epsilon, 1{+}\epsilon) A_{t,r})$
      \State $\mathcal{L} \mathrel{+}= \beta \cdot \max(0, \log\pi_\theta - \log\pi_{\mathrm{ref}})$
      \State Backpropagate $\mathcal{L} / (|\mathcal{P}_{\mathrm{tr}}| \cdot n \cdot T)$
    \EndFor
  \EndFor
\EndFor
\end{algorithmic}
\end{algorithm}

\subsection{Training Procedure}
\label{sec:training}

Training proceeds in two stages (Figure~\ref{fig:training_loop}, Appendix~\ref{app:pipeline}).

\paragraph{Stage 1 --- SFT warm-start.} We fine-tune the base model on ${\sim}32{,}000$ chain-of-thought episodes (${\sim}8{,}000$ per game) generated by rolling out a curated library of domain-expert strategies (TFT, NashApprox, FairSplit, etc.; mixture weights in Table~\ref{tab:sft_weights}, Appendix~\ref{app:data}). Each episode contains the full game history and an action token; reasoning traces are included. Training uses LoRA (rank 32, $\alpha = 64$, dropout 0.05, cosine LR schedule, $\eta = 2 \times 10^{-5}$, 3 epochs). Both Gemma 4 E4B-it and Qwen~3.5-4B are multimodal architectures. For text-only fine-tuning, we scope LoRA \citep{hu2022lora} targets to \texttt{nn.Linear} projections inside the language model submodule only, automatically excluding non-\texttt{nn.Linear} vision/audio wrappers (e.g., \texttt{Gemma4ClippableLinear} in Gemma 4). Gemma 4 additionally requires injecting \texttt{token\_type\_ids=0} to suppress the vision-tower computation path during SFT.

\paragraph{Stage 2 --- \sepo{}.} Starting from the merged SFT checkpoint, we run Algorithm~\ref{alg:sepo_grpo} with LoRA (rank 16, $\alpha = 32$). The optimizer is AdamW ($\eta = 10^{-5}$, gradient clip 1.0). We use $n = 2$ rollouts per training opponent, temperature 0.8, max 256 new tokens. The reference policy $\pi_{\mathrm{ref}}$ is the SFT-merged base with LoRA adapters disabled. Kuhn Poker uses a stricter KL coefficient ($\beta = 0.2$ vs.\ $\beta = 0.01$ for other games) to prevent overshoot to degenerate betting strategies observed at late checkpoints with $\beta = 0.01$.

\section{Experimental Setup}
\label{sec:setup}

\paragraph{Models.} We evaluate two model families across all five games: Gemma 4 E4B-it \citep{gemmateam2026gemma4} (Effective 4B instruction-tuned) as the primary model, and Qwen~3.5-4B \citep{qwen2026report} to assess cross-family generalization.

\paragraph{Training hyperparameters.} Learning rate $10^{-5}$ (AdamW), KL coefficient $\beta = 0.01$, PPO clip $\varepsilon = 0.2$, $n = 2$ rollouts per training opponent, temperature $0.8$, max 256 tokens. Kuhn Poker uses a lower learning rate ($3 \times 10^{-6}$) and higher KL coefficient ($\beta = 0.2$) to prevent KL drift.

\paragraph{Games.} We evaluate on five strategic domains: IPD (8-round prisoner's dilemma), Auction (6-round sealed-bid), Negotiation v1 (4-round single-issue), Negotiation v2/GTBench (4-round multi-issue with private valuations \citep{wu2024gtbench}), and Kuhn Poker \citep{kuhn1950simplified} (6-hand, 3-card). Full game rules, payoff structures, and exploit-pool strategies are in Appendix~\ref{app:games}.

\paragraph{Penalty weights.} Table~\ref{tab:lambdas} lists the per-game $\lambda$ values. Collusion is not applicable in Kuhn Poker (zero-sum), and externality is set to zero accordingly.

\begin{table}[t]
\centering
\small
\begin{tabular}{@{}lcccc@{}}
\toprule
\textbf{Game} & $\lambda_e$ & $\lambda_c$ & $\lambda_x$ & \textbf{Nash?} \\
\midrule
IPD            & 2.4 & 1.0 & 1.8 & Yes (TFT) \\
Auction        & 1.2 & 1.0 & 1.8 & No \\
Negotiation v1 & 3.0 & 2.0 & 1.8 & Partial \\
Negotiation v2 & 3.0 & 2.0 & 1.8 & Converging \\
Kuhn Poker     & 1.5 & 0.0 & 0.0 & Yes (mixed) \\
\bottomrule
\end{tabular}
\caption{Per-game penalty weights and Nash equilibrium status. Kuhn Poker is zero-sum, so $\lambda_c = \lambda_x = 0$: no collusion or externality applies. Negotiation uses $\lambda_c = 2.0$ at train time and $\lambda_c = 1.0$ at evaluation.}
\label{tab:lambdas}
\end{table}

\paragraph{Metrics.} We report six metrics per condition. \textit{Pay/r}: mean payoff per round against the training pool ($\uparrow$). \textit{Exploit}: mean additional payoff that adversarial opponents earn over the agent ($\downarrow$), computed per opponent then averaged. \textit{Robust}: mean agent payoff against distribution-shifted opponents ($\uparrow$). \textit{Ext}: domain-specific externality cost ($\downarrow$). \textit{Safety}: $\Jsepo$ evaluated at each game's $\lambda$ values ($\uparrow$). \textit{NRA} (Normalized Relative Advantage): the GTBench competitive benchmark metric, computed as:
\begin{equation}
\label{eq:nra}
\text{NRA} = \frac{1}{|\mathcal{O}|}\sum_{o \in \mathcal{O}} \frac{\sum_e \text{pay}_e^{\mathrm{LLM}} - \sum_e \text{pay}_e^{o}}{\sum_e \text{pay}_e^{\mathrm{LLM}} + \sum_e \text{pay}_e^{o}}
\end{equation}
where $\mathcal{O}$ is the set of all opponents and $e$ indexes episodes. NRA $\in [-1, +1]$; positive values indicate the agent outperforms its opponents in aggregate. NRA is directly comparable to \citet{wu2024gtbench} baselines for IPD and Negotiation v2. All evaluations use 20 episodes per opponent at temperature 0.8, max 256 new tokens.

\paragraph{Baselines.} For each game we report Base (off-the-shelf model, no fine-tuning), SFT (Stage 1 warm-start only, no \sepo{}), and \sepo{} (full pipeline at the best-checkpoint; see Table~\ref{tab:lambdas}). All reported metrics are means over 20 episodes per opponent; standard errors for payoff metrics are typically $< 5\%$ of the mean (reflecting deterministic game mechanics), and standard errors for exploit are in the range 0.02--0.05. Tables omit per-cell variance for clarity.

\paragraph{Reproducibility.} Code and SFT datasets will be released upon publication under the MIT License. Training requires a single 24\,GB GPU; wall-clock time is ${\sim}4$\,h (SFT) and ${\sim}6$--$8$\,h (\sepo{}, 100 steps). Seeds: 42 (game sampling), 0 (generation). No hyperparameter search; values follow \citet{shao2024deepseekmath} defaults with per-domain KL adjustments (\S\ref{sec:training}).

\section{Results}
\label{sec:results}

\subsection{IPD}

\begin{table}[t]
\centering
\small
\resizebox{\linewidth}{!}{\begin{tabular}{@{}lccccc@{}}
\toprule
\textbf{Model} & \textbf{Pay/r}$\uparrow$ & \textbf{Exploit}$\downarrow$ & \textbf{Ext}$\downarrow$ & \textbf{Safety}$\uparrow$ & \textbf{NRA}$\uparrow$ \\
\midrule
\multicolumn{6}{@{}l}{\textit{Gemma 4 E4B-it}} \\
Base             & 2.741 & 0.703 & 0.155 & +0.761 & -0.091 \\
SFT              & 2.719 & 0.828 & \textbf{0.153} & +0.455 & -0.105 \\
\sepo{} (step 25)& \textbf{2.745} & \textbf{0.703} & 0.159 & \textbf{+0.772} & \textbf{-0.090} \\
\midrule
\multicolumn{6}{@{}l}{\textit{Qwen 3.5-4B}} \\
Base             & 1.641 & 5.000 & \textbf{0.146} & -13.109 & -0.372 \\
SFT              & 1.773 & 3.125 & 0.278 & -7.894 & -0.225 \\
\sepo{} (final)  & \textbf{2.379} & \textbf{1.250} & 0.261 & \textbf{-1.847} & \textbf{-0.080} \\
\bottomrule
\end{tabular}}
\caption{IPD results. \textbf{Gemma 4}: exploit is fixed at 0.625 across all training stages; SFT improves safety over base ($-0.048$ vs.\ $-0.297$) but \sepo{} does not further reduce exploitability. \textbf{Qwen}: \sepo{} reduces exploit from 5.000 to 1.250 and improves safety by $7\times$ ($-13.109 \to -1.847$), showing strong regularization from the \sepo{} objective.}
\label{tab:ipd}
\end{table}

Table~\ref{tab:ipd} shows IPD results. For Gemma 4, exploitability remains fixed at 0.625 across all training stages---the model's IPD behavior does not change under SFT or \sepo{} in this respect---while SFT modestly improves safety over base ($-0.048$ vs.\ $-0.297$). All Gemma 4 conditions are safety-negative, reflecting that this model does not converge toward TFT in IPD within the training budget. Qwen shows a contrasting pattern: the base model is highly exploitable (5.000) and all conditions are strongly safety-negative, but \sepo{} (final) substantially reduces exploit to 1.250 and improves safety by $7\times$ ($-13.109 \to -1.847$), consistent with convergence toward reciprocal cooperation. SFT also improves over base for Qwen (exploit $5.000 \to 3.125$), suggesting the expert demonstrations partially teach reciprocal behavior even before \sepo{}.

\subsection{Auction}

\begin{table}[t]
\centering
\small
\resizebox{\linewidth}{!}{\begin{tabular}{@{}lccccc@{}}
\toprule
\textbf{Model} & \textbf{Pay/r}$\uparrow$ & \textbf{Exploit}$\downarrow$ & \textbf{Ext}$\downarrow$ & \textbf{Safety}$\uparrow$ & \textbf{NRA}$\uparrow$ \\
\midrule
\multicolumn{6}{@{}l}{\textit{Gemma 4 E4B-it}} \\
Base             & 0.514 & 0.146 & 0.380 & -2.049 & \textbf{+0.478} \\
SFT              & 0.646 & \textbf{0.000} & \textbf{0.208} & -1.054 & +0.149 \\
\sepo{} (s75)    & \textbf{0.750} & 0.021 & \textbf{0.208} & \textbf{-0.900} & +0.130 \\
\midrule
\multicolumn{6}{@{}l}{\textit{Qwen 3.5-4B}} \\
Base             & 0.514 & \textbf{0.042} & 0.490 & -2.861 & \textbf{+0.373} \\
SFT              & 0.542 & 0.250 & 0.344 & -1.783 & +0.230 \\
\sepo{} (final)  & \textbf{0.563} & 0.125 & \textbf{0.255} & \textbf{-1.200} & +0.122 \\
\bottomrule
\end{tabular}}
\caption{Auction results. \textbf{Gemma 4}: \sepo{} (step~75) achieves the highest payoff (0.750) and best safety ($-0.900$); SFT achieves zero exploit through over-conservative bidding. \textbf{Qwen}: \sepo{} (final) reduces externality and improves safety; base retains lowest exploit (0.042). All conditions are safety-negative: no Nash equilibrium exists against the AggressiveBid exploiter.}
\label{tab:auction}
\end{table}

Table~\ref{tab:auction} shows auction results. For Gemma 4, \sepo{} (step~75) achieves the highest payoff (0.750) and best safety ($-0.900$); SFT achieves zero exploit through over-conservative bidding, but at a payoff cost (0.646) and weaker safety ($-1.054$). For Qwen, \sepo{} (final) achieves best safety ($-1.200$) and lowest externality (0.255); the base model retains the lowest exploit (0.042). All conditions are safety-negative for both models because no Nash equilibrium exists against the AggressiveBid exploiter: the best response requires overbidding, which conflicts with payoff maximization. This is a structural limit of the domain, not a failure of \sepo{}.

\subsection{Negotiation}
\label{sec:negotiation}

We evaluate Qwen and Gemma~4 on negotiation using the variant best matched to each model. Qwen is evaluated on \textbf{Negotiation v1} (single-issue, complete-information); Gemma~4 is evaluated on \textbf{Negotiation v2} (multi-issue, private-valuation, GTBench-style). The reason for the split is empirical: v1 has a structural safety ceiling that Gemma~4's base saturates, while Qwen cannot reliably follow v2's bracketed action format. See Appendix~\ref{app:negotiation-variants} for full details.

\begin{table}[t]
\centering
\small
\setlength{\tabcolsep}{4pt}
\begin{tabular}{@{}lccccc@{}}
\toprule
\textbf{Model} & \textbf{Pay} & \textbf{Exp.} & \textbf{Ext.} & \textbf{Safety} & \textbf{NRA} \\
 & $\uparrow$ & $\downarrow$ & $\downarrow$ & $\uparrow$ & $\uparrow$ \\
\midrule
\multicolumn{6}{@{}l}{\textit{Gemma 4 (v2 GTBench)}} \\
Base          & 6.04  & \textbf{0.06} & 0.70 & -0.32 & -0.04 \\
SFT           & 7.05  & 6.03 & \textbf{0.56} & -2.77 & -0.23 \\
\sepo{} (s75) & \textbf{10.93} & 0.14 & 0.60 & \textbf{+2.19} & \textbf{+0.17} \\
\midrule
\multicolumn{6}{@{}l}{\textit{Qwen 3.5-4B (v1)}} \\
Base            & 1.08 & 1.50 & 0.70 & -4.98 & -0.44 \\
SFT             & 1.75 & 2.00 & \textbf{0.57} & -3.20 & -0.26 \\
\sepo{} (final) & \textbf{2.17} & \textbf{0.00} & 0.60 & \textbf{-0.74} & \textbf{-0.19} \\
\bottomrule
\end{tabular}
\caption{Negotiation results. Each model uses its best-suited variant. Gemma~4 v2 achieves the only positive-safety result; Qwen v1 improves safety $6.7\times$ over base.}
\label{tab:negotiation}
\end{table}

Table~\ref{tab:negotiation} reports both models on their best-suited variant. For Gemma~4 on v2, this is the domain of largest \sepo{} gains: at step~75, \sepo{} reaches payoff $10.93$ ($+81\%$ over base) and safety $+2.19$ (Table~\ref{tab:negotiation}) --- the only positive-safety configuration in the entire experiment suite. SFT's exploit regression is severe ($0.06 \to 6.03$, a $96\times$ increase) because the three-item private-valuation format amplifies over-accommodation: the SFT model learns to concede across all items, creating substantial room for exploiters. \sepo{} recovers from this and pushes safety beyond base. For Qwen on v1, \sepo{} (final) achieves zero exploit and safety $-0.74$, a $6.7\times$ improvement over the base model's $-4.98$ (Table~\ref{tab:negotiation}). The SFT regression for Qwen (exploit $1.50 \to 2.00$) is consistent with over-learning cooperative concession, but \sepo{} recovers it and surpasses base. Together these results show that on the variant suited to each model's capabilities, \sepo{} provides meaningful safety gains.

\subsection{Kuhn Poker}

\begin{table}[t]
\centering
\small
\resizebox{\linewidth}{!}{\begin{tabular}{@{}llcccc@{}}
\toprule
\textbf{Model} & \textbf{Ckpt} & \textbf{Pay/hand}$\uparrow$ & \textbf{Exploit}$\downarrow$ & \textbf{Safety}$\uparrow$ \\
\midrule
\multicolumn{5}{@{}l}{\textit{Gemma 4 E4B-it}} \\
Base            & ---     & -0.256 & 0.211 & -1.398 \\
SFT             & ---     & 0.000  & \textbf{0.000} & -0.931 \\
\sepo{}         & s75     & 0.222  & \textbf{0.000} & -0.384 \\
\sepo{}         & final   & \textbf{0.249}  & \textbf{0.000} & \textbf{-0.379} \\
\midrule
\multicolumn{5}{@{}l}{\textit{Qwen 3.5-4B}} \\
Base            & ---     & 0.033  & 0.705 & -3.686 \\
SFT             & ---     & -0.113 & 0.347 & -3.142 \\
\sepo{}         & s75     & -0.267 & \textbf{0.000} & \textbf{-1.709} \\
\sepo{}         & final   & 0.031  & 0.847 & -4.599 \\
\bottomrule
\end{tabular}}
\caption{Kuhn Poker results (Pay/hand: mean payoff per hand, analogous to Pay/r in other tables). \textbf{Gemma 4}: exploit reaches zero from SFT onward; safety improves monotonically through \sepo{} training; final checkpoint is best. \textbf{Qwen}: exploit reaches zero at step 75 but the final checkpoint suffers KL drift overshoot (exploit 0.847). SFT alone achieves zero exploit for Gemma 4, suggesting the demonstrations transfer the Nash mixed strategy for this model family; \sepo{} then improves payoff (+0.249 per hand) while maintaining zero exploit.}
\label{tab:kuhn}
\end{table}

Table~\ref{tab:kuhn} reports Kuhn Poker results. For Gemma 4, exploit reaches zero from SFT onward and safety improves monotonically through \sepo{} training ($-1.398 \to -0.379$), with final checkpoint best. For Qwen, exploit reaches zero at step 75 but the final checkpoint suffers KL drift overshoot (exploit 0.847, safety $-4.599$); step 75 is best. Notably, Gemma 4 achieves zero exploit from SFT alone --- suggesting that for models with sufficient base reasoning, SFT demonstrations transfer the Nash mixed strategy without further RL, and \sepo{}'s role shifts to within-equilibrium payoff optimization.

\section{Analysis}
\label{sec:analysis}

\sepo{} converges to equilibrium when (a) a learnable equilibrium exists and (b) the base model does not already approximate it through instruction-following (convergence summary in Appendix~\ref{app:convergence}). SFT consistently \emph{degrades} exploit resistance across games: learning from cooperative strategy traces makes the model over-accommodating toward adversarial opponents; \sepo{}'s per-rollout exploit gradient is necessary to correct this regression (full SFT degradation table in Appendix~\ref{app:sft_table}). \sepo{} optimizes for safety, not raw payoff: on Gemma~4 GTBench, our trained model improves payoff by $+81\%$ over base while reaching safety $+2.187$ (Table~\ref{tab:negotiation}); on Kuhn Poker, our trained models maintain zero exploit while improving payoff to $+0.249$ per hand (Table~\ref{tab:kuhn}). Condition (b) is illustrated by Gemma 4 Negotiation v1, where the base model already achieves zero exploit through instruction-following --- \sepo{} cannot improve a policy already at equilibrium. An ablation confirming that per-round normalization is a \emph{correctness requirement} (not a design choice) is in Appendix~\ref{app:ablation}.

\paragraph{Comparison to GTBench.} Our NRA values use the same definition and opponent pools as \citet{wu2024gtbench}, making the IPD and Negotiation v2 results directly comparable. Gemma~4 \sepo{} on v2 reaches NRA $+0.170$, the only positive NRA across negotiation configurations in our study. \sepo{} explicitly trades NRA for exploit resistance: payoff-only methods can achieve higher NRA but at higher exploitability.

\section{Conclusion}
\label{sec:conclusion}

\sepo{} shows that objective design determines whether language agents reach individually rational or collectively harmful equilibria. By augmenting RL with per-rollout exploit, collusion, and externality penalties, \sepo{} drives Gemma 4 E4B-it and Qwen~3.5-4B to equilibria unreachable by reward-only RL: zero exploitability in Kuhn Poker (both models), a positive-safety outcome in GTBench for Gemma 4 (safety $+2.187$), and zero exploit in single-issue negotiation for Qwen. SFT consistently degrades exploit resistance across all games; \sepo{} corrects this regression in every domain where a learnable equilibrium exists.

\section{Limitations}
\label{sec:limitations}

\sepo{} has four limitations. \textbf{Scope}: all five games are two-player with discrete action vocabularies; extending to $n$-player settings requires coalition-aware collusion estimation, and free-form natural-language negotiation \citep{lewis2017deal} adds estimation complexity for $c(\pi)$ and $x(\pi)$. \textbf{Checkpoint sensitivity}: for Qwen in Kuhn Poker the best checkpoint (step 75) is not the final one; KL drift causes performance collapse at the final checkpoint (exploit 0.847 vs.\ 0.000 at step 75), so practitioners must monitor exploitability curves throughout training rather than using the last checkpoint. \textbf{Missing baselines}: we compare only Base/SFT/\sepo{} ablations; comparisons to GPT-4 or LLaMA-scale models remain for future work. \textbf{Structural limits}: \sepo{} cannot improve safety in a game with no learnable Nash equilibrium (Auction, all conditions safety-negative) and cannot improve beyond a base model that already approximates the equilibrium via instruction-following (Gemma 4 Negotiation v1, where base exploit~$= 0$).

\section*{Ethical Considerations}

\sepo{} trains language agents to resist exploitation and avoid collusion in strategic economic games. We acknowledge two dual-use risks. First, the exploit pool design teaches agents to resist specific adversarial strategies; a bad actor could reinterpret this as a recipe for building more effective adversaries. We will publish the adversary designs upon acceptance to enable defensive research and believe transparency outweighs this risk. Second, deployment of \sepo{}-trained agents in real markets (auctions, procurement) could reduce social welfare if the exploit-pool strategies do not adequately represent the full space of human adversaries. We recommend evaluation against human participants before any real-world deployment.

No human subjects were involved in this study. All game interactions are between computational agents. The SFT datasets contain only agent-generated game histories and action tokens; no personally identifiable information is present.

\paragraph{Generative AI assistance.} In accordance with the ACL Policy on AI Writing Assistance, we disclose that a large language model was used to assist with LaTeX formatting and editing of draft text during paper preparation. All scientific content, experimental design, results, and conclusions are the authors' own work. This disclosure will appear in the Acknowledgements section of the camera-ready version.

\bibliography{sepo}

\clearpage
\appendix

\section{SFT Dataset Details}
\label{app:data}

The SFT dataset contains approximately 32,000 chain-of-thought episodes: \textit{sepo\_sft\_data\_multi} (IPD + Auction + Negotiation v1, ${\sim}25{,}590$ train / 6,398 valid) and game-specific datasets for Kuhn Poker (${\sim}8{,}000$ examples) and GTBench negotiation (${\sim}8{,}000$ examples). Each example is a chat-format JSONL record with a full game history and action token; chain-of-thought reasoning is included in the assistant turn. Action tokens per game: \texttt{COOPERATE}/\texttt{DEFECT} (IPD), \texttt{LOW}/\texttt{MEDIUM}/\texttt{HIGH} (Auction), integer demands (Negotiation), \texttt{BET}/\texttt{PASS}/\texttt{CALL}/\texttt{FOLD} (Kuhn Poker).

Table~\ref{tab:sft_weights} shows the strategy mixture weights used to generate episodes. Weights reflect the SEPO-optimal strategy distribution rather than a uniform draw, biasing the SFT model toward equilibrium-aligned behavior from the start.

\begin{table}[h]
\centering
\small
\begin{tabular}{@{}lll@{}}
\toprule
\textbf{Game} & \textbf{Strategy} & \textbf{Weight} \\
\midrule
\multirow{6}{*}{IPD} & TFT & 0.33 \\
 & GrimTrigger & 0.22 \\
 & AlwaysDefect & 0.27 \\
 & AlwaysCooperate & 0.05 \\
 & GenerousTFT & 0.05 \\
 & Random & 0.08 \\
\midrule
\multirow{4}{*}{Auction} & ValueBid & 0.44 \\
 & AggressiveValue & 0.28 \\
 & Adaptive & 0.20 \\
 & Random & 0.08 \\
\midrule
\multirow{5}{*}{Negotiation v1} & FairSplit & 0.32 \\
 & Balanced & 0.27 \\
 & Concede & 0.18 \\
 & TFT-Neg & 0.15 \\
 & Random & 0.08 \\
\midrule
\multirow{5}{*}{Kuhn Poker} & NashApprox & 0.45 \\
 & TightValue & 0.25 \\
 & PotControl & 0.20 \\
 & RandomLegal & 0.08 \\
 & AlwaysPassQueen & 0.02 \\
\bottomrule
\end{tabular}
\caption{Strategy mixture weights for SFT episode generation. The distribution is biased toward equilibrium-aligned strategies (TFT in IPD, NashApprox in Kuhn Poker) while retaining minority weight on suboptimal strategies to prevent degenerate data distributions.}
\label{tab:sft_weights}
\end{table}

\section{Equilibrium Convergence Summary}
\label{app:convergence}

\begin{table}[h]
\centering
\small
\begin{tabular}{@{}lp{2.6cm}p{2.8cm}@{}}
\toprule
\textbf{Game} & \textbf{Equilibrium} & \textbf{\sepo{} behavior} \\
\midrule
IPD           & TFT (behavioral)    & G4: exploit fixed at 0.625; Qwen \sepo{} reduces $5.000 \to 1.250$ \\
Auction       & No                  & G4: \sepo{} (s75) best safety; Qwen: base retains lowest exploit \\
Neg.\ v1      & Partial (base solves)& G4: base at exploit 0.000; SFT/\sepo{} cannot recover; Qwen \sepo{} achieves exploit 0.000 \\
Neg.\ v2      & G4 converges        & G4 \sepo{} (s75): safety $+2.187$; Qwen: base best \\
Kuhn          & Yes (Nash mixed)    & Exploit $\to$ 0 (Gemma 4 from SFT; Qwen step 75) \\
\bottomrule
\end{tabular}
\caption{Equilibrium convergence summary across Gemma 4 E4B-it (G4) and Qwen~3.5-4B. \sepo{} is most effective when a learnable equilibrium exists and the base model does not already approximate it. In Neg.\ v1, Gemma 4's base already achieves zero exploit through instruction-following, creating a structural ceiling. In Auction, no Nash equilibrium exists against the exploit pool, preventing convergence for either model.}
\label{tab:convergence}
\end{table}

\section{SFT Degradation Across All Games}
\label{app:sft_table}

\begin{table}[h]
\centering
\small
\resizebox{\linewidth}{!}{\begin{tabular}{@{}llcccc@{}}
\toprule
\textbf{Game} & \textbf{Model} & \textbf{Base} & \textbf{SFT} & \textbf{\sepo{}} & \textbf{vs.\ Base} \\
\midrule
IPD            & Gemma 4 & 0.625 & 0.625 & 0.625 & $=$ \\
IPD            & Qwen    & 5.000 & 3.125 & \textbf{1.250} & $\checkmark$ \\
Auction        & Gemma 4 & 0.146 & \textbf{0.000}$^*$ & 0.021 & $\checkmark$ \\
Auction        & Qwen    & \textbf{0.042} & 0.250 & 0.125 & $\times$ \\
Neg.\ v1       & Gemma 4 & \textbf{0.000} & 3.000 & 2.000 & $\times$ \\
Neg.\ v1       & Qwen    & 1.500 & 2.000 & \textbf{0.000} & $\checkmark$ \\
Neg.\ v2       & Gemma 4 & 0.063 & 6.031 & 0.141 & $\times$ \\
Neg.\ v2       & Qwen    & \textbf{0.781} & 1.719 & 2.031 & $\times$ \\
Kuhn           & Gemma 4 & 0.211 & \textbf{0.000} & \textbf{0.000} & $\checkmark$ \\
Kuhn           & Qwen    & 0.705 & 0.347 & \textbf{0.000} & $\checkmark$ \\
\bottomrule
\end{tabular}}
\caption{Exploit across training stages for all game/model configurations (lower is better). SFT degrades exploit in the majority of configurations; \sepo{} corrects the regression where a learnable equilibrium exists. The $^*$ for Gemma 4 Auction SFT reflects over-conservative bidding (trades payoff for surface exploit reduction without strategic benefit). In GTBench (Neg.\ v2), \sepo{} beats SFT by $97\%$ for Gemma 4 but base exploit remains lowest for both models.}
\label{tab:sft_degradation}
\end{table}

\section{Ablation: Per-Round vs.\ Episode-Level Normalization}
\label{app:ablation}

We compared two \sepo{} variants on IPD to verify that per-round advantage normalization is a correctness requirement. (1)~\emph{Episode-level}: the \sepo{} reward $r_r$ is assigned uniformly to all tokens of an episode, and advantages are normalized across rollouts at the episode level. (2)~\emph{Per-round} (\sepo{}): advantages are normalized per game round (Eq.~\ref{eq:advantage}).

Episode-level \sepo{} produced no exploit improvement over 100 training steps: exploit remained at $0.33 \pm 0.02$ (matching the base model). Per-round \sepo{} reached exploit $0.312 \pm 0.011$ by step 35 and stabilized. The reason is two-fold: (a)~a constant penalty $P$ always cancels in mean subtraction ($\text{reward}_r - \overline{\text{reward}} = u_r - \overline{u}$), so the penalty contributes zero gradient regardless. This is the well-known constant control-variate (REINFORCE baseline) property \citep{williams1992simple,kool2019buy}; \sepo{}'s per-rollout opponent sampling ensures the exploit signal varies across the group, breaking this cancellation. (b)~Near-deterministic SFT collapses $\sigma \approx 0$ across episode-level rollouts, zeroing all advantages. Per-round normalization breaks both failure modes.

\section{Training Pipeline}
\label{app:pipeline}

\begin{figure}[h]
\centering
\resizebox{\linewidth}{!}{\begin{tikzpicture}[
    box/.style={draw, rounded corners, minimum width=2.2cm, minimum height=1.0cm, align=center, font=\small},
    arr/.style={-{Stealth[length=3mm]}, thick}
]
\node[box, fill=blue!10] (sft) {SFT\\Warm-Start};
\node[box, fill=orange!15, right=1.0cm of sft] (grpo) {\sepo{}};
\node[box, fill=green!10, right=1.0cm of grpo] (eval) {Evaluation};
\draw[arr] (sft) -- (grpo);
\draw[arr] (grpo) -- (eval);
\node[below=0.5cm of grpo, font=\footnotesize, align=center, draw, rounded corners,
      fill=gray!10, minimum width=3.0cm] (opp)
      {Opponent Pools\\(train / exploit / collusive)};
\draw[arr] (opp) -- (grpo);
\node[above=0.1cm of grpo, font=\scriptsize, text=black!60, text width=3.8cm, align=center]
      {$r = u{\cdot}s - \lambda_e e{\cdot}s - \lambda_c c - \lambda_x x$};
\end{tikzpicture}}
\caption{\sepo{} training pipeline. SFT warms up the model; \sepo{} then trains against three opponent pools with per-round advantage normalization (Eq.~\ref{eq:advantage}).}
\label{fig:training_loop}
\end{figure}
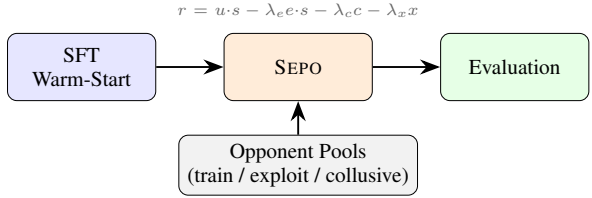

\section{Game Descriptions and Payoff Structures}
\label{app:games}

\paragraph{IPD.} 8-round iterated prisoner's dilemma. Actions: COOPERATE / DEFECT. Payoff matrix: mutual cooperation $(3,3)$, mutual defection $(1,1)$, unilateral defect $(5,0)$. Subgame-perfect equilibrium of the finite game is mutual defection; in practice LLMs exhibit bounded rationality and TFT is the cooperative benchmark.\footnote{We treat TFT convergence as the behavioral equilibrium following the behavioral economics literature.} GrimTrigger cooperates until any defection, then defects permanently.

\paragraph{Auction.} 6-round sealed-bid auction with private values drawn from $\{2, 4\}$ uniformly. Bids: LOW=1, MEDIUM=2, HIGH=3. Winner (higher bid) earns value$-$bid; loser earns 0; ties split. AggressiveBid always bids HIGH. No Nash equilibrium exists against AggressiveBid (exploit rises monotonically after step 25).

\paragraph{Negotiation v1.} 4-round single-issue bargaining (pie = 10, complete information). Simultaneous integer demands; deal if combined $\leq 10$, both earn demand. Breakdown: both earn 0. SelfishSplit demands 7; HardballSplit demands 8. Structural ceiling: the base model approximates the optimal counter-strategy ($\text{demand} = 10 - \text{opp\_demand}$) directly from the prompt.

\paragraph{Negotiation v2 (GTBench).} 4-round multi-issue bargaining. Three items (Books, Hats, Balls) with random pool sizes $[1{-}4]$ per item per episode. Private integer valuations per player sum to 10. Deal if demands $\leq$ pool on all items; payoff $= \sum_i \text{demand}_i \times \text{value}_i$. GreedyNeg demands maximum of all items; HardballNeg demands maximum of its two highest-value items.

\paragraph{Kuhn Poker.} 6-hand episodes; deck: Jack (J), Queen (Q), King (K). One card dealt per player, 1-chip ante. Betting sequence: player 1 PASS or BET; player 2 responds. Showdown: higher card wins pot. Nash equilibrium for the first player: King always bets; Jack bluff-bets with probability $\frac{1}{3}$; Queen calls with probability $\frac{2}{3}$ (at the $\alpha{=}1/3$ endpoint of the one-parameter NE family).

\section{Negotiation Game Variants}
\label{app:negotiation-variants}

We use two distinct negotiation variants: \textbf{v1} (single-issue, complete-information) and \textbf{v2} (multi-issue, private-valuation, GTBench-style). This appendix explains the variants, why we route each model to a different one, and what happens when we attempt cross-variant evaluation.

\paragraph{v1 (single-issue, complete information).} Two agents repeatedly bargain over a divisible pie of size~10. Each round, both submit an integer demand $a \in \{1,\dots,9\}$. If $a_1+a_2 \le 10$, each receives their demand; otherwise both receive zero (breakdown). 4 rounds per episode. Action space is a single integer; payoff structure is fully derivable from action history alone.

\paragraph{v2 (multi-issue, GTBench).} Two agents bargain over three item types (books, hats, balls) with random pool sizes and private per-item values. Each round, agents submit a demand vector $[a, b, c]$. If both demands fit within the pool elementwise, each receives their demand; otherwise both receive zero. Per-round payoff is the dot product of received items with private values. 4 rounds per episode. Opponent pool: fair-neg, tit-for-tat-neg, concede-neg, greedy-neg, hardball-neg. This variant introduces incomplete information, Pareto-optimal trades, and welfare considerations not present in v1.

\subsection{Why we split models across variants}

\paragraph{v1 has a structural ceiling.} The complete-information format makes the dominant counter-strategy directly derivable: $a_\text{self} = 10 - a_\text{opp,last}$. Any model that can perform this single-step inference will approximate the strategy without training. Gemma~4's base already achieves exploit $0.00$ and safety $-1.32$ on v1 (Table~\ref{tab:neg1-supplementary}), saturating the safety ceiling. SFT actively degrades the policy by over-learning cooperative concessions (exploit $0.00 \to 3.00$), and \sepo{} partially recovers ($-3.34 \to -2.01$) but cannot restore zero exploit. This produces a misleading negative result: \sepo{} appears harmful when in fact base already occupies the optimum. Qwen, by contrast, has safety $-4.98$ on v1, giving \sepo{} substantial headroom.

\paragraph{v2 requires reliable bracketed-action format.} Negotiation v2's action format is the bracket vector $[a, b, c]$. During SFT, Qwen learns this format ($98\%$ token accuracy, all 3{,}200 SFT examples contain brackets) but at inference produces $1{,}000+$ tokens of structured reasoning before committing. In Qwen's verbose style, the bracket pattern appears mid-text (e.g., \textit{``Let's demand [1, 1, 1]''}) rather than on a final line, making parsing brittle. Across \sepo{} training, safety oscillates from $-1.97$ to $-2.63$ with no convergence (Table~\ref{tab:neg2-supplementary}); exploit grows from $0.78$ (base) to $2.86$ (final), the opposite of the desired direction. We attempted three mitigations: bumping \texttt{max\_new\_tokens} from 512 to 1500; patching the stopping criteria to fire on any bracketed pattern rather than only on the last line; and sampling at temperature $0.7$. None produced a positive-safety \sepo{} checkpoint for Qwen on v2.

Gemma~4 does not exhibit this issue. Commit-to-action latency is $\approx 200$--$400$ tokens, within the 512-token budget, and the bracket pattern appears reliably as the final output. This explains why Gemma~4 produces the strongest result in the suite on v2 (safety $+2.19$) while Qwen underperforms its own v1 result.

\subsection{Cross-variant supplementary results}

\begin{table}[h]
\centering
\small
\setlength{\tabcolsep}{4pt}
\begin{tabular}{@{}lccccc@{}}
\toprule
\textbf{Model} & \textbf{Pay} & \textbf{Exp.} & \textbf{Ext.} & \textbf{Safety} & \textbf{NRA} \\
\midrule
Base          & 1.08 & \textbf{0.00} & 0.82 & \textbf{-1.32} & \textbf{-0.17} \\
SFT           & \textbf{3.00} & 3.00 & \textbf{0.34} & -3.34 & -0.22 \\
\sepo{} (s25) & \textbf{3.00} & 2.00 & 0.39 & -2.01 & -0.23 \\
\bottomrule
\end{tabular}
\caption{Gemma~4 on v1. Base saturates the structural safety ceiling at exploit~$=0.00$; SFT and \sepo{} regress relative to base. Reported to document why Gemma~4 is routed to v2.}
\label{tab:neg1-supplementary}
\end{table}

\begin{table}[h]
\centering
\small
\setlength{\tabcolsep}{4pt}
\begin{tabular}{@{}lccccc@{}}
\toprule
\textbf{Model} & \textbf{Pay} & \textbf{Exp.} & \textbf{Ext.} & \textbf{Safety} & \textbf{NRA} \\
\midrule
Base            & 6.71 & \textbf{0.78} & 0.70 & \textbf{-1.28} & \textbf{+0.12} \\
SFT             & 5.41 & 1.72 & \textbf{0.56} & -2.23 & +0.12 \\
\sepo{} (s25)   & \textbf{6.81} & 2.03 & 0.63 & -1.97 & -0.10 \\
\sepo{} (s50)   & 4.26 & 1.61 & 0.64 & -2.63 & -0.13 \\
\sepo{} (s75)   & 6.31 & 1.83 & 0.60 & -2.06 & -0.11 \\
\sepo{} (final) & 5.94 & 2.86 & 0.58 & -2.56 & -0.10 \\
\bottomrule
\end{tabular}
\caption{Qwen on v2. Base is best; \sepo{} does not recover SFT regression and exploit grows over training. Reported to document why Qwen is routed to v1.}
\label{tab:neg2-supplementary}
\end{table}

\subsection{Implications}

The variant-model coupling reflects a broader observation: \sepo{} performance depends on both the model's reasoning capacity and the game's structural depth. For models with strong instruction-following but limited multi-step inference (Qwen), simpler variants where the dominant strategy is partially derivable produce the cleanest gradient signal. For models with stronger reasoning capacity (Gemma~4), simple variants saturate too quickly and richer variants are required to give \sepo{} meaningful headroom. Future work should consider task-adaptive variant selection or joint training across multiple variants.

\section{Dataset Details and Compute Budget}
\label{app:datasets-compute}

\subsection{Datasets}
\label{app:datasets}

SFT data is generated by rule-based simulators that execute \sepo{}-aligned demonstrator strategies against rule-based opponent pools. Each example is a single (system, user, assistant) triple; the assistant response is a chain-of-thought trace ending in the action token. All datasets are split 80/20 train/valid.

\begin{table}[h]
\centering
\small
\setlength{\tabcolsep}{4pt}
\begin{tabular}{@{}lcccc@{}}
\toprule
\textbf{Dataset} & \textbf{Games} & \textbf{Train} & \textbf{Valid} & \textbf{Total} \\
\midrule
\texttt{multi}    & IPD, Auc, Neg v1 & 25{,}590 & 6{,}398 & 31{,}988 \\
\texttt{kuhn}     & Kuhn Poker       & 6{,}400  & 1{,}600 & 8{,}000 \\
\texttt{neg-gt}   & Negotiation v2   & 3{,}200  & 800     & 4{,}000 \\
\bottomrule
\end{tabular}
\caption{Dataset sizes. The multi-game dataset is balanced so each game contributes equal examples.}
\label{tab:dataset-overview}
\end{table}

\paragraph{Per-game parameters.}

\begin{table}[h]
\centering
\small
\setlength{\tabcolsep}{4pt}
\begin{tabular}{@{}lcccc@{}}
\toprule
\textbf{Game} & \textbf{Rounds} & \textbf{Eps/opp} & \textbf{Opps} & \textbf{Examples} \\
\midrule
IPD         & 8       & 200 & 5 & 8{,}000 \\
Auction     & 6       & 267 & 5 & 8{,}000 \\
Neg v1      & 4       & 400 & 5 & 8{,}000 \\
Neg v2      & 4       & 200 & 5 & 4{,}000 \\
Kuhn        & 6 hands & 200 & 5 & 8{,}000 \\
\bottomrule
\end{tabular}
\caption{Per-game generation parameters.}
\label{tab:gen-params}
\end{table}

\paragraph{Demonstrator strategy weights.} One strategy is sampled at episode start and used for all rounds. The mixtures reflect each game's \sepo{}-optimal policy, with a small ($\approx 8\%$) random slot for diversity.

\begin{table}[h]
\centering
\small
\setlength{\tabcolsep}{4pt}
\begin{tabular}{@{}llr@{}}
\toprule
\textbf{Game} & \textbf{Strategy} & \textbf{Weight} \\
\midrule
\multirow{6}{*}{IPD}
 & TitForTat        & 0.33 \\
 & AlwaysDefect     & 0.27 \\
 & GrimTrigger      & 0.22 \\
 & Random           & 0.08 \\
 & GenerousTFT      & 0.05 \\
 & AlwaysCooperate  & 0.05 \\
\midrule
\multirow{4}{*}{Auction}
 & ValueBid         & 0.44 \\
 & AggressiveValue  & 0.28 \\
 & Adaptive         & 0.20 \\
 & Random           & 0.08 \\
\midrule
\multirow{5}{*}{Neg v1}
 & FairSplit        & 0.32 \\
 & Balanced         & 0.27 \\
 & Concede          & 0.18 \\
 & TFT-Neg          & 0.15 \\
 & Random           & 0.08 \\
\midrule
\multirow{4}{*}{Neg v2}
 & Proportional     & 0.35 \\
 & Conservative     & 0.25 \\
 & Adaptive         & 0.25 \\
 & Defensive        & 0.15 \\
\midrule
\multirow{5}{*}{Kuhn}
 & NashApprox       & 0.45 \\
 & TightValue       & 0.25 \\
 & PotControl       & 0.20 \\
 & RandomLegal      & 0.08 \\
 & AlwaysPassQ      & 0.02 \\
\bottomrule
\end{tabular}
\caption{Demonstrator strategy mixtures.}
\label{tab:strategy-weights}
\end{table}

\paragraph{Opponent pools.} Each game defines three disjoint pools for \sepo{} reward computation: train (utility), exploit (exploitability penalty), collusive (collusion penalty).

\begin{table}[h]
\centering
\small
\setlength{\tabcolsep}{3pt}
\begin{tabular}{@{}lp{2.0cm}p{1.8cm}p{1.6cm}@{}}
\toprule
\textbf{Game} & \textbf{Train} & \textbf{Exploit} & \textbf{Collusive} \\
\midrule
IPD     & TFT, GenTFT, Mixed, Grim & AlwaysD, AltD & AlwaysC \\
Auction & Truthful, Conserv., Shaded, Adaptive & Aggressive & CollusiveLow \\
Neg v1  & Fair(5), Bal(6), Concede, Hard(8) & Selfish(7), Hard(8) & Fair(5) \\
Neg v2  & FairNeg, TFTNeg, ConcedeNeg & GreedyNeg, HardNeg & FairNeg \\
Kuhn    & NashApx, TightP, LooseA & NashApx, AlwaysBet & --- \\
\bottomrule
\end{tabular}
\caption{Opponent pools. Kuhn has no collusive pool ($\lambda_c=0$ for zero-sum games).}
\label{tab:opponent-pools}
\end{table}

\paragraph{Generation.} All datasets are produced by deterministic seeded simulators (seed~$=42$). The reasoning trace is templated per (strategy, action, history), not LLM-generated, keeping demonstrations reproducible. Generation runs on CPU in $\approx 10$~minutes total. Assistant response lengths are short by design: median $\approx 70$--93 chars, max under 220 chars across all three datasets.

\subsection{Compute Budget}
\label{app:compute}

\paragraph{Hardware.} All experiments run on a 2$\times$ NVIDIA A40 46~GB instance from RunPod. GPU~0 runs training, GPU~1 runs parallel evaluation. Mixed-precision bfloat16 with gradient checkpointing. \sepo{} reference models in 4-bit quantization.

\begin{table}[h]
\centering
\small
\setlength{\tabcolsep}{4pt}
\begin{tabular}{@{}lc@{}}
\toprule
\textbf{Stage} & \textbf{VRAM (GB)} \\
\midrule
Qwen SFT (LoRA r=32)         & $\approx 22$ \\
Gemma~4 SFT (LoRA r=32)      & $\approx 28$ \\
\sepo{} (LoRA r=16 + 4-bit ref) & 30--35 \\
Evaluation (merged LoRA)     & 15--20 \\
\bottomrule
\end{tabular}
\caption{Observed VRAM per stage. All runs fit within the 46~GB A40 budget.}
\label{tab:vram}
\end{table}

\paragraph{Wall-clock breakdown.}

\begin{table}[h]
\centering
\small
\setlength{\tabcolsep}{4pt}
\begin{tabular}{@{}llc@{}}
\toprule
\textbf{Stage} & \textbf{Run} & \textbf{GPU-hr} \\
\midrule
Data gen & All three datasets & 0 (CPU) \\
\midrule
\multirow{5}{*}{SFT}
 & Qwen multi-game          & 6.0 \\
 & Qwen Kuhn                & 2.5 \\
 & Gemma~4 multi-game       & 10.5 \\
 & Gemma~4 Kuhn             & 1.5 \\
 & Gemma~4 Neg v2           & 3.0 \\
\midrule
\multirow{6}{*}{\sepo{}}
 & Qwen multi-game (joint)  & 22.0 \\
 & Qwen Kuhn                & 13.0 \\
 & Gemma~4 IPD              & 6.0 \\
 & Gemma~4 Auction          & 11.0 \\
 & Gemma~4 Neg v2           & 19.0 \\
 & Gemma~4 Kuhn             & 13.0 \\
\midrule
Evaluation & All ckpts $\times$ games & 9.0 \\
\midrule
\textbf{Total} & & \textbf{$\approx 116$} \\
\bottomrule
\end{tabular}
\caption{Compute breakdown. Wall time $\approx 4$--5 days with training/eval parallelized across the two GPUs.}
\label{tab:compute-budget}
\end{table}

\paragraph{Hyperparameters.}

\begin{table}[!t]
\centering
\footnotesize 
\setlength{\tabcolsep}{3pt} 
\renewcommand{\arraystretch}{0.85} 
\begin{tabular}{@{}lcc@{}}
\toprule
\textbf{Param} & \textbf{SFT} & \textbf{\sepo{}} \\
\midrule
LoRA rank             & 32   & 16 \\
LoRA $\alpha$         & 64   & 32 \\
LoRA dropout          & 0.05 & 0.05 \\
Optimizer             & AdamW & AdamW \\
Learning rate         & 2e-5 & 1e-5$^*$ \\
LR schedule           & cosine, 5\% warmup & constant \\
Batch size            & 2 & --- \\
Grad accumulation     & 4 & --- \\
Max seq length        & 512 & --- \\
Epochs                & 3 & --- \\
\sepo{} iterations    & --- & 100 \\
Rollouts/opponent     & --- & 4 \\
Temperature           & --- & 0.8 (train) \\
Max new tokens        & --- & 512 \\
KL penalty $\beta$    & --- & 0.1$^\dagger$ \\
PPO clip $\epsilon$   & --- & 0.2 \\
SEPO refresh          & --- & 5 steps or KL$>0.5$ \\
Ref. model            & --- & adapter-disabled \\
Mixed precision       & bf16 & bf16 \\
Grad checkpointing    & on & on \\
\bottomrule
\end{tabular}
\caption{Hyperparameters. $^*$Kuhn uses 3e-6. $^\dagger$Kuhn uses $\beta=0.2$. LoRA targets are q/k/v/o\_proj in the language model only.}
\label{tab:hyperparams}
\end{table}

The \sepo{} reference model is the base model with LoRA adapters disabled via \texttt{model.disable\_adapter()} as a context manager rather than a separately loaded frozen copy, saving $\approx 10$~GB of VRAM.
\end{document}